# Independence of spin-orbit torques from the exchange bias direction in $Ni_{81}Fe_{19}$/IrMn bilayers

Hilal Saglam[1, 2, a], J. Carlos Rojas-Sanchez[3], Sebastien Petit[3], Michel Hehn[3], Wei Zhang[4], John E. Pearson[1], Stephane Mangin[3], Axel Hoffmann[1]

[1] Materials Science Division, Argonne National Laboratory, Lemont IL 60439, USA

[2] Department of Physics, Illinois Institute of Technology, Chicago IL 60616, USA

[3] Université de Lorraine, CNRS, IJL, F-54000 Nancy, France

[4] Department of Physics, Oakland University, Rochester MI 48309, USA

a) saglam@anl.gov

**Abstract**

We investigated a possible correlation between spin Hall angles and exchange bias in $Ni_{81}Fe_{19}$/IrMn samples by performing spin torque ferromagnetic resonance measurements. This correlation is probed by patterning of $Ni_{81}Fe_{19}$/IrMn bilayers in different relative orientations with respect to the exchange bias direction. The measured voltage spectra allow a quantitative determination of spin Hall angles, which are independent of the orientation around 2.8±0.3%.

Recently, antiferromagnetic materials have received increased interest in spintronics devices[1] beyond their traditional use in exchange-biased based applications.[2,3] Antiferromagnets are magnetically ordered with zero net-magnetization, which makes them insensitive to external magnetic fields. For the same reason, they do not produce any magnetic stray fields, which avoids cross-coupling between different devices in close proximity. Furthermore, they have very high characteristic frequencies in the terahertz regime; hence they can operate at high speeds. The discovery of several magneto-transport effects in antiferromagnets such as spin-orbit torques,[4] anisotropic magneto resistance,[5,6,7] spin Seebeck,[8,9] inverse spin Hall,[10,11] Galvanic effects,[12] and efficient spin current transmission[13,14,15] show the feasibility of using them as active components for spintronic devices. In particular, the discovery of electrical switching and readout of an antiferromagnet by spin–orbit torque impressively shows that antiferromagnets can be controlled electrically in similar ways as their ferromagnetic counterparts.[16]

Of particular interest is the possibility to drive the magnetization dynamics of a ferromagnet using spin the Hall effect,[17,18,19,20] which generates a spin accumulation on the lateral surface of a charge carrying material. The efficiency of the spin Hall effect is described by the spin Hall angle,[21] $\Theta_{SH} = \frac{2e}{\hbar}(J_s/J_c)$. Using the spin torque ferromagnetic resonance technique,[22] spin Hall effects in metallic CuAu I-type antiferromagnets (IrMn, FeMn, PtMn, and PdMn) have been investigated and spin Hall angles are reported between 0.02-0.08.[9,10] It has been also shown that spin orbit torques in PtMn are sufficiently strong for switching the magnetization of an adjacent Co/Pt layer.[23] Theoretically, it is expected that the spin current injection from an antiferromagnet to a normal metal varies as the antiferromagets' texture changes.[24] These studies indicate that the interface plays an important role for the spin current injection and transmission.

Another interfacial effect is exchange bias and its origin is traditionally accepted to be due to exchange coupling between a ferromagnet and an antiferromagnet at the interface.[25] This interfacial interaction has been extensively used in many spintronics devices such as spin valves, magnetic tunnel junctions where it provides a reference magnetization orientation.[26,27] By using exchange bias, the direction of antiferromagnetic order at the interface can be manipulated at a microscopic level. Thus, the natural question to ask is whether there is a direct correlation between exchange bias and spin-orbit torques as the ferromagnetic/antiferromagnetic interface is crucial for both phenomena.

Previous studies intentionally avoided exchange bias by adding a Cu layer between the ferromagnetic and antiferromagnetic layer. While this Cu-layer eliminates direct magnetic interfacial coupling, it is highly transparent for spin currents. Here, we removed the Cu layer sandwiched between ferromagnet and antiferromagnet and performed spin torque ferromagnetic resonance measurement in $Ni_{81}Fe_{19}$/IrMn bilayers with different relative orientations of applied fields and exchange bias. As it will be discussed below, these measurements revealed no significant dependence of spin-orbit torques on the exchange bias field directions.

Multiple Ni$_{81}$Fe$_{19}$/IrMn bilayers with lateral dimension of 20 µm* 90 µm were fabricated on SiO$_2$ substrate using magnetron sputtering, photolithography and ion milling based on our previous recipe[28]. Subsequently, contact pads [Ti (3 nm)/Au (150 nm)] were deposited on the samples for electrical measurements, see Fig. 1 (a). In order to establish exchange bias, the fabricated antiferromagnet/ferromagnet bilayers were annealed at 250 ºC and subsequently cooled down to room temperature in the presence of 600 Oe magnetic field. Note that all the devices with different relative orientations were fabricated on the same chip and annealed simultaneously.

Spin transfer torque ferromagnetic resonance measurements were performed for all the samples with various relative orientations between exchange bias and external field [Fig. 1 (b)]. A schematic of the measurement setup is illustrated in Fig. 1(a). An oscillating charge current was driven through *rf*-probes in contact with the contact pads. This oscillating charge current generates a transverse spin current, which exerts a torque on magnetization of Ni$_{81}$Fe$_{19}$. The *rf* current flowing through the Ni$_{81}$Fe$_{19}$ layer mixes with the time-dependent resistivity of the Ni$_{81}$Fe$_{19}$ due to anisotropic magnetoresistance in response to the oscillating magnetization, which results in a rectified *dc* voltage that was measured as a function of applied magnetic field. During the measurements, a 10-mW fixed microwave current was applied with frequencies varied from 8 to 15 GHz and the external magnetic field was kept constant at 45º with respect to the *rf* current direction [Fig. 1(c)].

Fig. 2 (a) shows typical measured spin torque ferromagnetic resonance signals for a Ni$_{81}$Fe$_{19}$/IrMn bilayer at frequencies of 8, 10, 12 and 14 GHz and θ=0º. As the microwave frequency increases, the resonance field shifts towards higher fields, which is in agreement with the Kittel model,[29] and the magnitude of measured *dc* voltage (*V*) decreases. Here, the measured *dc* voltages are sums of symmetric and antisymmetric Lorentzian components, which can be attributed to damping-like and field-like torques, respectively. Fig. 2 (b) shows a spin torque ferromagnetic spectra, which was measured at 8 GHz and decomposed into its symmetric (*V$_s$*) and antisymmetric (*V$_a$*) Lorentzian voltage contributions. In order to illustrate the angular dependence of exchange bias on spin orbit torques, we examine the ratio of symmetric to antisymmetric voltage contributions, as well as associated effective spin Hall angles[30] for both negative and positive external fields. We extracted the symmetric and antisymmetric voltage contributions by fitting the spin torque ferromagnetic resonance curves:[31]

$$V = \left[V_S \frac{\Delta H^2}{\Delta H^2 + (H_{ext} - H_{res})^2} + V_A \frac{\Delta H(H_{ext} - H_{res})}{\Delta H^2 + (H_{ext} - H_{res})^2}\right], \quad (1)$$

where $\Delta H$ is resonance linewidth, $H_{ext}$ is the applied field, and $H_{res}$ is the resonance field for ferromagnetic resonance at a given frequency. The effective spin Hall angles can be quantified via the ratio of symmetric and antisymmetric voltage contributions:[20]

$$\Theta_{SH}^{eff} = \frac{V_s}{V_a} \frac{e\mu_0 M_s t_{AF} t_{FM}}{\hbar} \left[1 + \frac{4\pi M_s}{H_{ext}}\right]^{1/2}, \qquad (2)$$

where $\mu_0$ is the permeability in vacuum, $M_s$ is the saturation magnetization of $Ni_{81}Fe_{19}$ which can be extracted by fitting the resonance frequency as a function of $H_{res}$ by Kittel equation:[27]

$$f_{res} = \frac{\gamma}{2\pi}[(H_{res} + H_B)(H_{res} + H_B + 4\pi M_s)]^{1/2}, \qquad (3)$$

where $H_B$ is the effective exchange bias field.

In exchange-biased ferromagnetic/antiferromagnetics bilayers, the exchange anisotropy gives rise to a resonance shift and enhanced linewidth broadening compared to unbiased films.[32] Patterning samples in four different directions with respect to the unidirectional exchange bias anisotropy, i.e., θ=0, 45, 90, 135º, enables us to control the shift in resonance field during measurements. Fig. 3 (a) shows the measured *dc* voltage as a function of field at 12 GHz for θ=0. When the measurement field and exchange bias field are parallel (perpendicular) to each other, we measured the maximum (minimum) resonance field shift of 60 Oe. The summary of electrically measured exchange bias fields for all angles (θ= 0, 45, 90, 180º) is shown in Fig. 3(b).

We also measure effective damping due to the exchange bias, which can be extracted from frequency dependence of linewidth broadening[33] (*ΔH*) [Fig. 4(a)]. Almost identical *ΔH* was observed for all directions, see [Fig. 4(b)]. If we consider that at least part of the effective damping is related to spin pumping into the adjacent IrMn, then this suggests that the efficiency of spin current transmission at the interface is approximately identical for all directions, *i.e.*, independent from the exchange bias direction. Furthermore, the additional effective damping for our $Ni_{81}Fe_{19}$/IrMn sample is larger than the measured value in the previous study[10] where exchange coupling at the interface eliminated by inserting a thin Cu layer. This can be attributed to magnetic losses in the antiferromagnetic spin lattices, which is also directly exchange coupled to the ferromagnetic magnetization in the $Ni_{81}Fe_{19}$ and therefore can provide additional damping.

Fig. 5 (a) shows the ratio of symmetric and antisymmetric voltages for frequencies ranging from 8 to 15 GHz. No angular dependence of $V_s/V_a$ was observed, which is consistent with the effective damping data. These findings suggest that both damping-like (given by $V_s$) and Oersted field-like torques (given by $V_a$) are independent of the local spin structure at the $Ni_{81}Fe_{19}$/IrMn interface. However, we observed that $V_s/V_a$ for positive and negative fields differ slightly. A possible explanation for this difference could be a resonant heating of the samples. Together with the asymmetric structure of the sample layout [see Fig. 1 (a)], this may result in a lateral temperature difference. Ultimately, this difference may generate additional Seebeck voltage across the sample, which is due to the resonant heating has an asymmetric Lorentzian lineshape.

Ignoring this offset, the averaged effective spin Hall angles for positive and negative fields for varying microwave excitation frequencies from 8 to 15 GHz are shown in Fig. 5 (a). The effective spin Hall angles for different relative orientations between exchange anisotropy and external field show very small variation, and we estimate $\Theta_{SH}^{eff} = (2.8 \pm 0.3)\%$. We observed slightly smaller spin Hall angles compared to previous measurements[11] in the presence of Cu interlayer between IrMn and $Ni_{81}Fe_{19}$.

In a previous study, by spin torque ferromagnetic resonance measurements on (001) and (111) oriented $IrMn_3/Ni_{81}Fe_{19}$ samples, it has been shown that non-collinear antiferromagnet $IrMn_3$ has a large spin Hall angle and it is facet dependent.[34] Upon performing out of plane field annealing, it was shown that this spin Hall angle can increase drastically, while in plane annealing shows no difference as is also observed in our study. On the other hand, our observations are inconsistent with a previous spin orbit ferromagnetic measurement experiments on NiFe/IrMn system where a large enhancement of damping-like torques was measured arising from the antiferromagnetic order at the interface.[35]

We conclude that there is no dependence of the effective spin Hall angles on the relative orientation of the in-plane exchange bias with respect to current flow and thus the polarization direction of the concomitant spin currents and accumulations. However, there are still strong spin orbit torques even when the antiferromagnet is directly exchange coupled to the ferromagnet. We also observe a similar trend in ferromagnetic resonance data, where the angular dependence of effective damping is almost identical for all directions, which may suggest that the spin current transmission through the $Ni_{81}Fe_{19}$/IrMn interface is not affected by the local magnetic structure.

The work at Argonne National Laboratory including experimental design, magneto-optic measurements, data analysis and manuscript preparation was supported by the U.S. Department of Energy, Office of Science, Basic Energy Sciences, Materials Science and Engineering Division under Contract No. DE-AC02-06CH11357. Work at the Université de Loraine including sample preparation and transport measurements was suported by the French PIA project "Lorraine Université d'Excellence", reference ANR-15-IDEX-04-LUE. Lithographic patterning was carried out at the Micro and Nano technology Platform of Lorraine (MINALOR).

**Figures:**

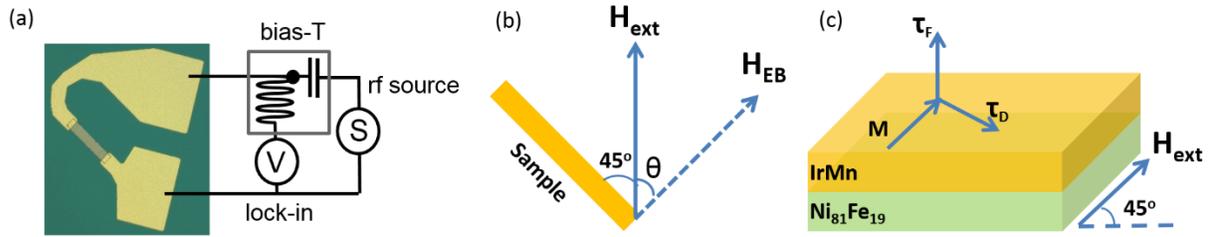

**Figure 1.** (a) Schematic representation of the spin torque ferromagnetic resonance experimental setup. (b) Relative orientations between exchange bias and external field during spin torque ferromagnetic resonance measurement. (c) Illustrative picture of a $Ni_{81}Fe_{19}/IrMn$ showing the spin transfer torques $\tau_F$, $\tau_D$, magnetization $M$, and external field $H_{ext}$.

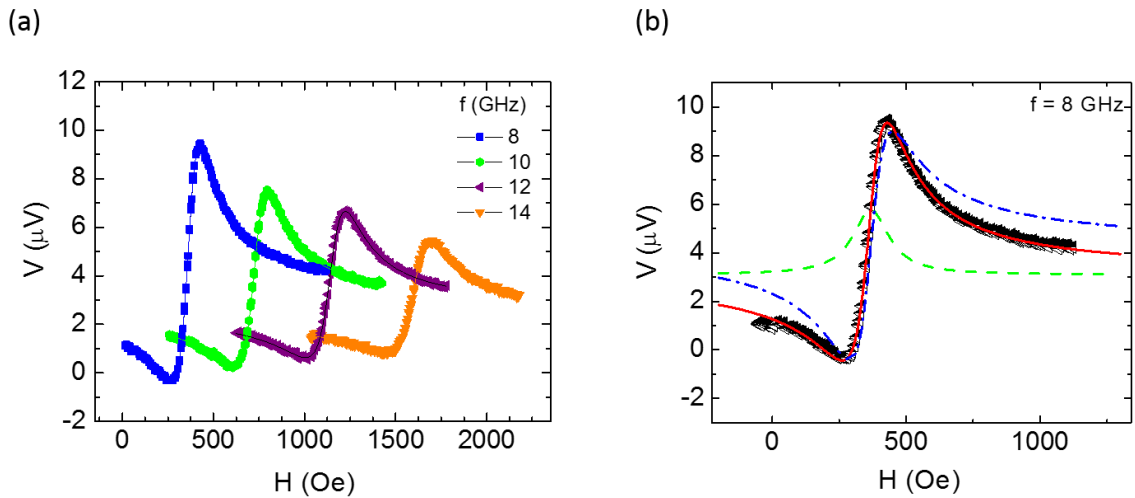

**Figure 2.** (a) Measured spin torque ferromagnetic resonance signals at 8, 10, 12 and 14 GHz with the angle θ=0° between the applied field and the exchange bias direction. (b) Single spin torque ferromagnetic resonance spectra measured at 8 GHz. The red solid line represents the fit to Lorentzian function. The green dashed and blue dash-dotted lines represent symmetric and antisymmetric voltage components, respectively.

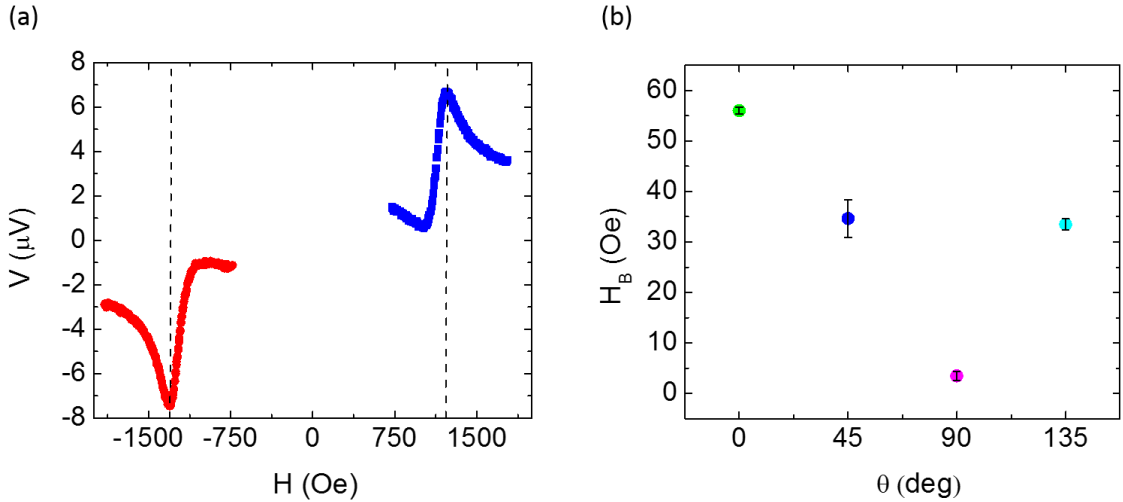

**Figure 3.** a) Spin torque ferromagnetic resonance spectra obtained at 12 GHz illustrate the resonance field shift due to exchange (for θ=0º). Blue and red curves depicts positive and negative field sweeps, respectively. (b) Measured in plane exchange bias field $H_B$, which represent the relative orientation of exchange bias anisotropy with respect to the measurement field $H_{ext}$.

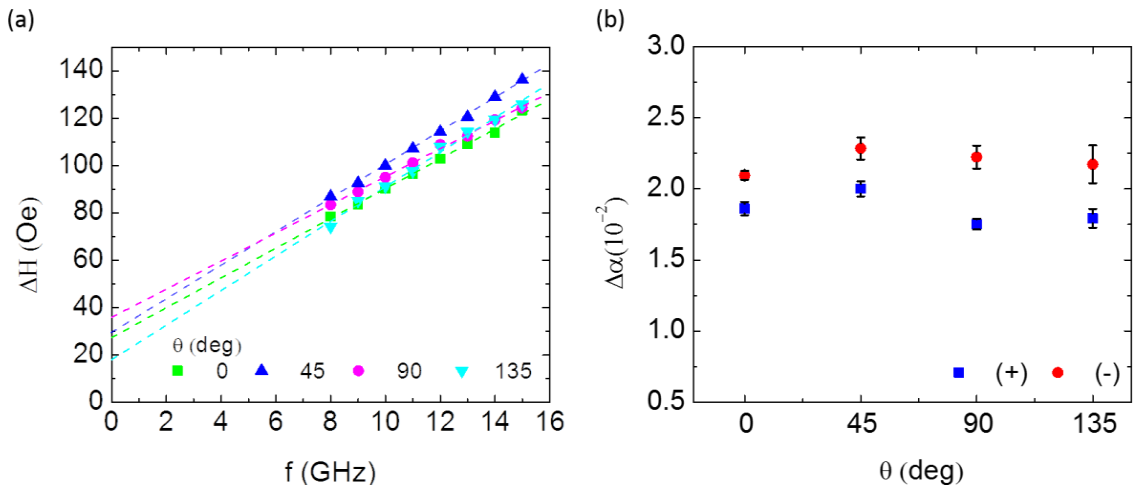

**Figure 4.** (a) Frequency dependence of the linewidth broadening $\varDelta H$, for varying $\theta$ (for positive field sweeps). (b) Effective damping $\varDelta\alpha$ as a function of $\theta$ for positive and negative fields.

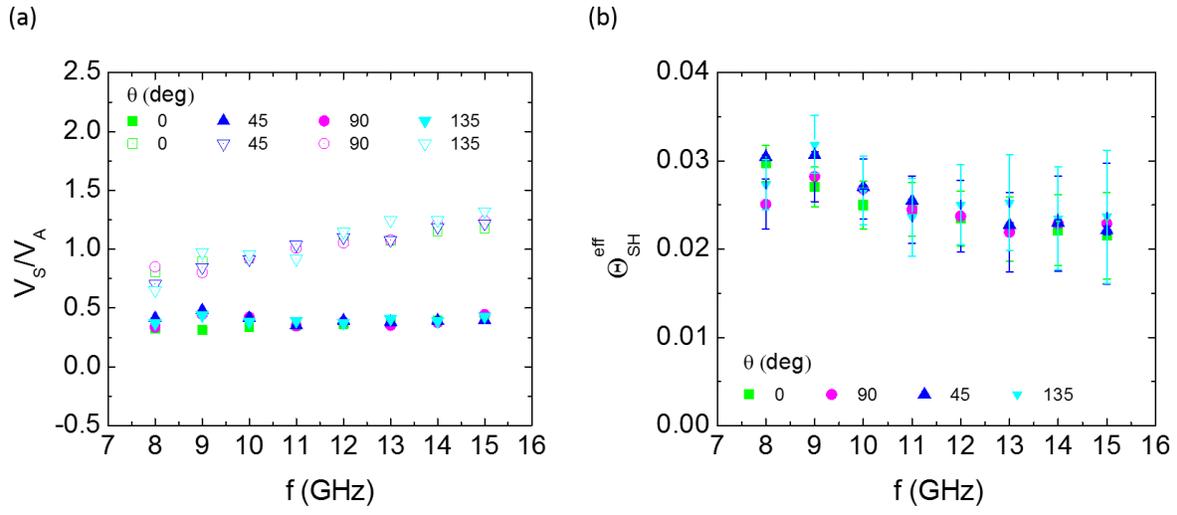

**Figure 5.** (a) Ratio of symmetric voltage to antisymmetric voltage for varying frequencies from 8 to 15 GHz. Open and close circles represent negative and positive field sweeping, respectively. (b) Effective spin Hall angles $\Theta_{SH}$ as a function of frequency for different $\theta$.

[1] A. Hoffmann and S. D. Bader, Phys. Rev. Appl. **4**, 047001 (2015).
[2] M. B. Jungfleisch, W. Zhang, and A. Hoffmann, Phys. Lett. A. **382**, 865 (2018).
[3] V. Baltz, A. Manchon, M. Tsoi, T. Moriyama, and Y. Tserkovnyak, Rev. Mod. Phys. **90**, 015005 (2018).
[4] J. Zelezny, H. Gao, K. Vyborny, J. Zemen, J. Masek, A. Manchon, J. Wunderlich, J. Sinova, and T. Jungwirth, Phys. Rev. Lett. **113**, 157201 (2014).
[5] B. G. Park, J. Wunderlich, X. Marti, V. Holy, Y. Kurosaki, M. Yamada, H. Yamamoto, A. Nishide, J. Hayakawa, H. Takahashi, A. B. Shick, and T. Jungwirth, Nat. Mater. **10**, 347 (2011).
[6] X. Marti, B. G. Park, J. Wunderlich, H. Reichlova, Y. Kurosaki, M. Yamada, H. Yamamoto, A. Nishide, J. Hayakawa, H. Takahashi, and T. Jungwirth, Phys. Rev. Lett. **108**, 017201 (2012).
[7] H. Wang, C. Du, P. C. Hammel, and F. Yang, Phys. Rev. Lett. **113**, 097202 (2014).
[8] S. M. Rezende, R. L. Rodríguez-Suárez, and A. Azevedo, Phys. Rev. B **93**, 014425 (2016).
[9] S. M. Wu, W. Zhang, A. KC, P. Borisov, J. E. Pearson, J. S. Jiang, D. Lederman, A. Hoffmann, and A. Bhattacharya, Phys. Rev. Lett. **116**, 097204 (2016).
[10] W. Zhang, M. B. Jungfleisch, W. Jiang, J. E. Pearson, A. Hoffmann, F. Freimuth, and Y. Mokrousov, Phys. Rev. Lett. **113**, 196602 (2014).
[11] W. Zhang, M. B. Jungfleisch, F. Freimuth, W. Jiang, J. Sklenar, J. E. Pearson, J. B. Ketterson, Y. Mokrousov, and A. Hoffmann, Phys. Rev. B **92**, 144405 (2015).
[12] J. Zelezny, H. Gao, K. Vyborny, J. Zemen, J. Masek, A. Manchon, J. Wunderlich, J. Sinova, and T. Jungwirth, Phys. Rev. Lett. **113**, 157201 (2014).
[13] H. Saglam, W. Zhang, M. B. Jungfleisch, J. Sklenar, J. E. Pearson, J. B. Ketterson, and A. Hoffmann, Phys. Rev. B **94**, 140412 (R) (2016).
[14] H. Wang, C. Du, P. C. Hammel, and F. Yang, Phys. Rev. Lett. **113**, 097202 (2014).
[15] L. Frangou, S. Oyarzún, S. Auffret, L. Vila, S. Gambarelli, and V. Baltz, Phys. Rev. Lett. **116**, 077203 (2016).
[16] P. Wadley, B. Howells, J. Železný, C. Andrews, V. Hills, R. P. Campion, V. Novák, K. Olejník, F. Maccherozzi, S. S. Dhesi, S. Y. Martin, T. Wagner, J. Wunderlich, F. Freimuth, Y. Mokrousov, J. Kuneš, J. S. Chauhan, M. J. Grzybowski, A. W. Rushforth, K. W. Edmonds, B. L. Gallagher, and T. Jungwirth, Science **351**, 587 (2016).
[17] M. I. D'yakonov and V. I. Perel, Sov. Phys. JETP Lett. **13**, 467 (1971).
[18] J. E. Hirsch, Phys. Rev. Lett. **83**, 1834 (1999).
[19] S. Zhang, Phys. Rev. Lett. **85**, 393 (2000).
[20] V. Tshitoyan, C. Ciccarelli, A. P. Mihai, M. Ali, A. C. Irvine, T. A. Moore, T. Jungwirth, and A. J. Ferguson, Phys. Rev. B **92**, 214406 (2015).
[21] A. Hoffmann, IEEE Trans. Magn. **49**, 5172 (2013).
[22] L. Liu, T. Moriyama, D.C. Ralph, and R.A. Buhrman, Phys. Rev. Lett. **106**, 036601 (2011).
[23] S. Fukami, C. Zhang, S. Dutta Gupta, A. Kurenkov, H. Ohno, Nat. Materials **15**, 535 (2016).
[24] H. B. M. Saidaoui, A. Manchon, and X. Waintal, Phys. Rev. B **89**, 174430 (2014).
[25] W. H. Meiklejohn and C. P. Bean, Phys. Rev. **105**, 904 (1957).
[26] J. Nogués, and I. K. Schuller, J. Magn. Magn. Mater. **192** (1999).
[27] A.H. MacDonald, and M. Tsoi, Philos. Trans. R. Soc. A **369**, 3098 (2011).
[28] G. Malinowski, M. Hehn, S. Robert, O. Lenoble, A. Schuhl, and P. Panissod, Phys. Rev. B **68**, 184404 (2003).
[29] C. Kittel, Phys. Rev. **73**, 155 (1948).

[30] J.-C. Rojas-Sánchez, N. Reyren, P. Laczkowski, W. Savero, J.-P. Attané, C. Deranlot, M. Jamet, J.-M. George, L. Vila and H. Jaffrès, Phys. Rev. Lett. **112**, 106602 (2014).

[31] L. Liu, C.-F. Pai, Y. Li, H. W. Tseng, D. C. Ralph, R. A. Buhrman, Science **336**, 555 (2012).

[32] M. Rubinstein, P. Lubitz, and S. Cheng, J. Magn. Magn. Mater. **195** (1999).

[33] T. L. Gilbert, IEEE Trans. Magn. Trans. **40**, 3443 (2004).

[34] W. Zhang, W. Han, S. Yang, Y. Sun, Y. Zhang, B. Yan and S. P. Parkin, Sci. Adv. **2**, 1600759 (2016).

[35] V. Tshitoyan, Phys. Rev. B **92**, 214406 (2015).